# Laser-Driven High-Velocity Microparticle Launcher In Atmosphere And Under Vacuum

David Veysset,[a,*] Yuchen Sun,[a,b] Steven E. Kooi,[a] Jet Lem,[a,b] and Keith A. Nelson.[a,b]

[a]Institute for Soldier Nanotechnologies, Massachusetts Institute of Technology, 77 Massachusetts Avenue, Cambridge, MA 02139, USA

[b]Department of Chemistry, Massachusetts Institute of Technology, 77 Massachusetts Avenue, Cambridge, MA 02139, USA

**Abstract**

This paper presents a novel approach to launch single microparticles at high velocities under low vacuum conditions. In an all-optical table-top method, microparticles with sizes ranging from a few microns to tens of microns are accelerated to supersonic velocities depending on the particle mass. The acceleration is performed through a laser ablation process and the particles are monitored in free space using an ultra-high-speed multi-frame camera with nanosecond time resolution. Under low vacuum, we evaluate the current platform performance by measuring particle velocities for a range of particle types and sizes, and demonstrate blast wave suppression and drag reduction under vacuum. Showing an impact on polyethylene, we demonstrate the capability of the experimental setup to study materials behavior under high-velocity impact. The present method is relevant to space applications, particularly to rendezvous missions where velocities range from tens of m/s to a few km/s, as well as to a wide range of terrestrial applications including impact bonding and impact-induced erosion.

**Highlights**

- A laser-driven microparticle launcher with vacuum capabilities is described
- Maximum velocities for particles of different types and masses are reported under atmospheric and low vacuum conditions
- Velocities are limited by optical absorption saturation and particle strength and heat resistance
- Under vacuum conditions, the blast wave generated upon laser ablation is largely suppressed and drag is reduced
- The ballistic limit is measured for a polyethylene film and the residual velocity is modeled using a Recht-Ipson model

Corresponding author. Tel.: +1-617-324-6425.
E-mail address: dveysset@mit.edu.

NOTATION
| | | | |
|---|---|---|---|
| $v$ | velocity of particle | $P_b$ | blast overpressure |
| $v_r$ | residual velocity of particle | $P_0$ | ambient pressure |
| $R$ | radius | $\gamma$ | specific heat ratio |
| $M_p$ | mass of particle | $c_0$ | speed of sound in ambient conditions |
| $A$ | area | $m_{pl}$ | mass of plug |
| $C_D$ | coefficient of drag | $v_{bl}$ | ballistic limit |
| $\rho$ | density | | |
| $t$ | time | Subscripts | |
| $E_l$ | laser energy | $p$ | particle |
| $E_0$ | laser threshold energy | $l$ | laser |
| $\eta$ | conversion efficiency | $m$ | medium |
| $M$ | Mach number | $b$ | blast |

## 1. Introduction

High velocity impact of microparticles is fundamental to many fields from additive manufacturing [1] and needleless drug delivery [2] to spacecraft protection against micro-debris and interstellar dust collection [3–6]. Dust sensors and collectors have been implemented in many missions to study high- and hyper-velocity dusts [6–9]. For dust sensors in particular, it has been challenging to isolate basic impact parameters such as mass, size, velocity, and density from *in-situ* recorded impact signals [10]; likewise, for dust collectors, it has been difficult to predict the penetration of dust and therefore to recover the precious particles [6]. Because these dusts are relevant to all missions as they pose a threat to spacecraft integrity and personnel [11], it is also important to design impact-resistant materials and architectures that will behave as expected under high and ultrahigh strain rate conditions in space and during atmospheric re-entry. Success in the design of such materials depends on relevant simulation work and on the development of ground facilities for impact testing capable of accessing the relevant size and velocity regimes.

Well-established facilities include vacuum free-fall tubes, light-gas guns (LGGs), powder guns, and Van de Graaf (VdG) accelerators. Each of these facilities each covers a specific ranges of particle size and impact speed. VdG accelerators (using nanoparticles and limited to conductive materials) produce multiple impacts per shot, making it difficult to isolate the response of a single particle [12]. Vacuum free-fall tubes enable single particle impacts but are limited to slow impacts, making them unsuitable for high-velocity applications. LGGs can typically accelerate larger particles in relatively large facilities [13]. There is currently no standard technique to generate microparticle impacts for sizes between a few microns and a hundred microns with velocities from tens of m/s to a few km/s.

Here, we present an in-house designed microparticle impact test as a complementary platform to existing methods that aims to produce high-velocity single microparticle impacts and bridge the current gap in ground-based instrumentation. The laser-induced particle impact test (LIPIT) is an all-optical, table-top platform that accelerates individual microparticles to high velocities and images each trajectory in real time using an ultra-high-speed camera [14]. LIPIT can accelerate particles with wide-ranging parameters such as material type (ceramic, metallic, polymeric, etc.), conductivity, dimensions, density, and shape, enabling systematic studies with many types of particles and substrates [14–16]. For instance, recent works from Hassani et al. focused on metallic particle impacts on metallic substrates to investigate impact bonding and impact-induced erosion mechanisms [17–19]. In other works, Veysset et al. and Hsieh et al. studied the high strain rate response of hierarchical elastomers to supersonic microparticle impact in order to elucidate energy dissipation mechanisms in this class of materials [20,21]. In this paper, we present a modified version of the LIPIT that allows for experimentation under vacuum and higher particle velocities than previously reported. We report for the first time the maximum velocities achievable using this method for a wide range of particle type, size, and mass, and discuss current limitations. We directly measure the drag acting on high-velocity microparticles by air in atmospheric conditions and examine the blast wave generated upon laser ablation. Finally, we demonstrate with impact on a polyethylene film how real-time observations of impact events can be used to infer materials properties under high-strain rate deformation.

## 2. Materials and methods

### 2.1. Microparticle launcher

The present method relies on laser ablation for acceleration, similar to laser-launched flyer plate methods [22–24]. As shown schematically in Fig. 1a, following laser ablation of the metallic film by a laser excitation pulse (pulsed Nd:YAG, 532-nm wavelength, 10-ns duration), microparticles are launched into free space between the launch pad and the target. The launch pad assembly is composed of a 210-µm-thick, 25-mm diameter, glass substrate (Corning No. 2 microscope cover slip) and a thin gold layer. The gold layer is deposited using e-beam evaporation to a desired thickness of 670 nm. Microparticles are then deposited on the substrates and spread on the metallic layer using lens cleaning papers and a drop of ethanol. Optionally, an elastomeric layer (30-µm polyurea, not shown in schematics) may be added onto the metallic film before deposition of particles as described, for instance, in Refs [17,25]. In the present case, particles reach higher velocities but are in direct contact with the ablation area. A pinhole with 75-µm diameter is inserted between the launch pad and the target, about 1 mm away from the launch pad, to block ablation-generated debris while allowing passage of the particle (see Fig. 1b). The pinhole is positioned in the path of the particle trajectory using a secondary CCD camera.

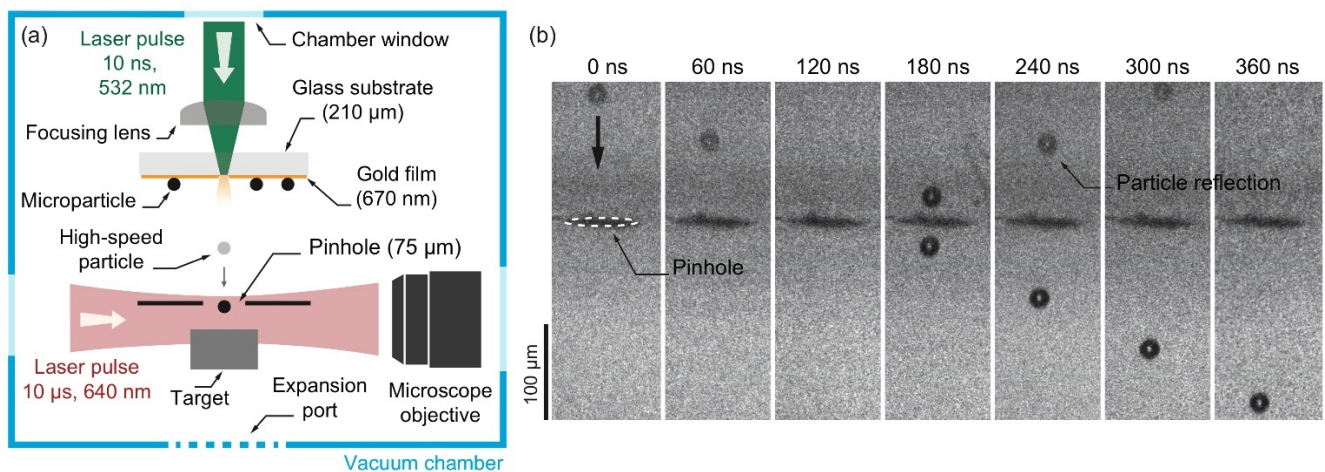

**Fig. 1.** (a) Microparticle launcher schematic. (b) Multi–frame sequence showing a 21-µm silica particle launched at 895 m/s going through the pinhole. The reflection of the particle from the pinhole holder is visible in the images. 7 of 16 frame are presented here, cropped from their initial size for ease of visualization. No target was present here. The time stamps, shown above the frames, indicate the delays relative to the first frame of the sequence.

The particle interaction with the target sample can be recorded in real time using an ultrahigh-frame-rate camera yielding 16 frames (SIMX16, Specialised Imaging) and a synchronized quasi-cw laser imaging pulse for illumination (Cavilux, Specialised Imaging 640-nm wavelength, 10-µs duration). The system can provide time resolution as short as 3 ns, enough to visualize a microparticle travelling up to 50 km/s. A wide range of particle velocities can be achieved (from 10 m/s to ~3 km/s) depending on the particle mass by tuning the energy of the laser excitation pulse. Moreover, because of the localized ablation by the laser excitation and the intentionally sparse distribution of the microparticles on the launch pad surface, a single particle is accelerated to high-velocity in each laser shot.

*2.2. Microparticles*

Silica particles were purchased from Thermo Fisher Scientific (three batches with 10 µm, 20 µm, and 30 µm nominal diameters) and from Microparticles GmbH (7.38-µm and 13.79-µm diameters). Steel particles were purchased from Cospheric (two batches with 10–25 µm and 27–38 µm diameters). Copper particles were purchased from Alfa Aesar (10-µm nominal diameter). Polystyrene particles were purchased from Microbeads (Dynoseeds TS10, 10-µm nominal diameter). Hollow glass particles were purchased from Cospheric (HGMS-0.67, 5–20 µm diameters). Preceding each shot, particles were selected using a secondary CCD camera and individual particle diameters were measured.

*2.3. Vacuum chamber design*

The vacuum chamber was designed to fit the launch pad, the focusing lens for laser excitation, the microscope objective, and the pinhole. It was purchased from Kurt J. Lesker Company with customized windows, feedthroughs, and ports. Windows include a 25-mm window (Brewster-angle cut) for laser pulse excitation and two 50-mm anti-reflection-coated windows for laser illumination and imaging optics. A 200-mm flange was added at the bottom of the chamber, in line with the particle impact direction, to enable the connection to an additional chamber for larger target accommodation (see expansion port in Fig. 1a). The internal dimensions of the chamber are 40×40×40 cm. The chamber is pumped with a rotary vane pump (Edwards, RV3). The minimum pressure attainable with the current design is $4\times10^{-3}$ mbar. The target can be placed at a desirable distance from the launch pad, from <1 mm to ~30 cm.

## 3. Platform performance and evaluation

*3.1. Velocity performance*

Figure 2a demonstrates how the laser pulse energy is adjusted to control particle velocity. We present the velocity dependence on laser energy for two different types of launch pad assemblies, the 'glass-gold' configuration (presented here) and the 'glass-gold-elastomer' configuration (previously reported [17]). In the absence of the elastomer layer, particles can be accelerated to higher velocities. For 7.4-µm silica particles, the maximum speeds reached with the 'glass-gold' configuration is 2950 m/s at a laser pulse energy of 6 mJ. On the other hand, with the 'glass-gold-elastomer' configuration, the maximum velocity is about 1050 m/s at laser energies above 6 mJ. In this configuration, the elastomer layer mitigates the ablative force accelerating the particles and reduces achievable velocities. However, while yielding lower velocities, the elastomer layer serves as containment for ablation debris and as a thermal shield between the ablation plume and the particle.

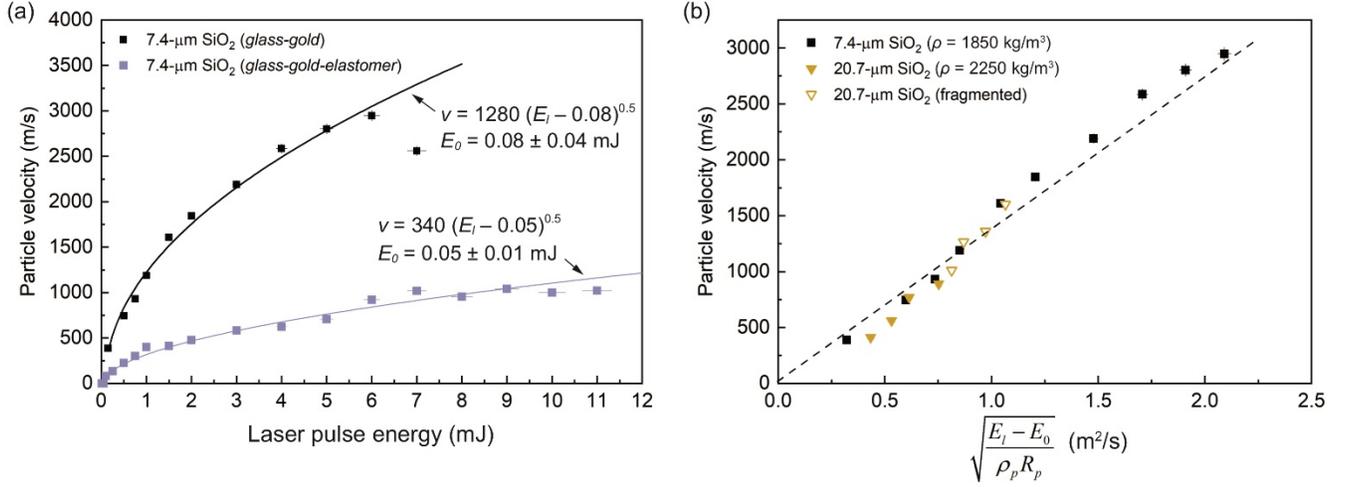

**Fig. 2.** (a) Particle velocity versus laser pulse energy for 7.4-μm silica particles for two launch pad configurations. Particle velocity saturates at laser energies higher than 6 mJ. (b) Maximum particle velocity versus laser energy over particle density and radius for two particles with different densities and masses. 20.7-μm silica particles fracture above a laser energy of 3 mJ. Open triangles mark velocities for silica fragments.

In the 'glass-gold' case, the particle velocity drops above a laser energy of 6 mJ due to the saturation in optical absorption of the metallic film and possible optical damage of the glass substrate. Likewise, in the 'glass-gold-elastomer' case, the velocity saturates above 6-7 mJ. Both curves are fitted with the following form:

$$v = \alpha(E_l - E_0)^\beta, \qquad (1)$$

where $\alpha$, $\beta$, and $E_0$ are fitting parameters. $E_0$ is the threshold energy that is necessary to detach the particle from the launch pad substrate. In both configurations, the power $\beta$ is found to be 0.5 and the threshold energies $E_0$ comparable (within uncertainty). Based on the obtained value for $\beta$, we hypothesize that the kinetic energy of the particle is a fraction of the deposited laser energy behind the particles. The laser is focused to a spot with a fixed diameter of 50 μm. The deposited total laser pulse energy is the total laser pulse energy $E_l$ normalized by the ratio between the particle cross section $A_p$ and the laser spot size $A_l$. $\eta$ is the conversion efficiency between laser energy and kinetic energy. This gives:

$$\tfrac{1}{2} M_p v^2 = \eta \frac{A_p}{A_l}(E_l - E_0), \qquad (2)$$

which yields

$$v = \sqrt{\tfrac{3}{2}\eta \frac{1}{A_l}} \sqrt{\frac{E_l - E_0}{\rho_p R_p}}. \qquad (3)$$

We examine the relation for two spherical particles with different densities, $\rho_p$, and radii, $R_p$ (Fig. 2b). $\eta$ is found to be 16% using the 7.4-μm particle data for fitting. The reasonable overlap between the velocity data for both particles validates our hypothesis.

Limitations in the launch pad and the particle dictate the currently obtainable velocities. For the launch pad, optical saturation of the laser absorption prevents further velocity increase past a certain laser energy, typically around 6 mJ as shown in Fig. 2a. For the particle, the sudden acceleration and rapid heating from laser ablation may induce fracture, melting or both under the substrate saturation energy. For instance, as illustrated in Fig. 3a, a 13.8-μm silica particle fragmented well under optical saturation at a laser energy of 3.0 mJ. In another example, Fig. 3b shows, a 50-μm aluminum particle that was melted and fragmented following laser ablation. This is detrimental when intact particles are desired, but could be used to study fragmented or molten particle impacts.

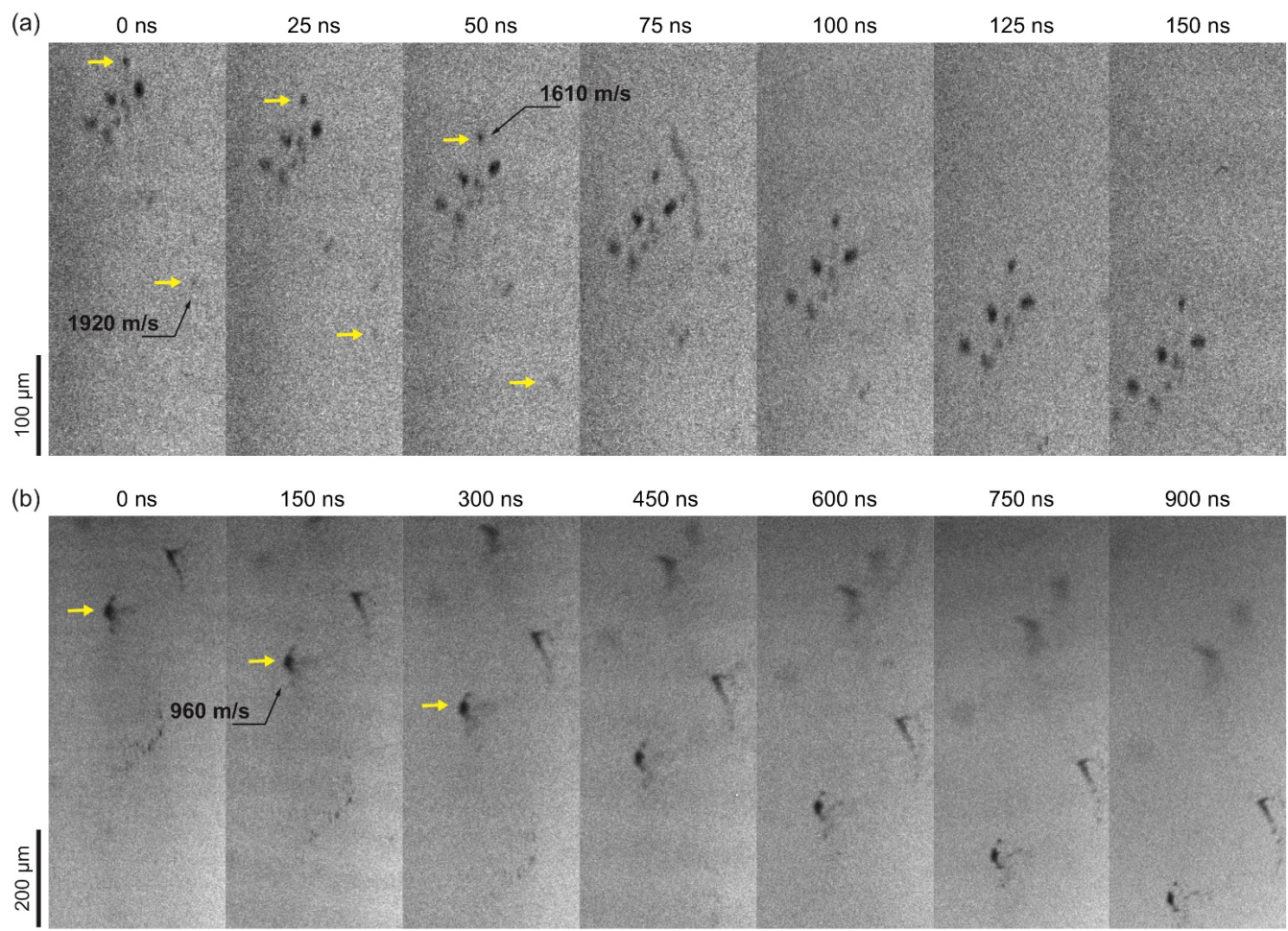

**Fig. 3.** Examples of damaged particles. (a) A 13.8-μm silica fragmented following laser ablation of the gold film, while (b) a 50-μm aluminum particle melted and fragmented.

Maximum velocities, for visibly intact particles, were measured for a wide range of particle masses both under atmospheric and vacuum conditions (Fig. 4). We note no difference in maximum particle velocity in either condition. We find an empirical power law between the maximum particle velocity and the particle mass. This law can help predict the maximum velocity for a given particle type and size and one can imagine that submicron particles could be accelerated to tens of km/s relevant to hypervelocity applications. However, such particles would fall below our optical resolution (~1 μm), and at such velocities would be blurred in a 3-ns measurement window, so direct velocity measurement would be difficult. With further improvements of the launch pad assembly and imaging conditions, we foresee in the near future higher achievable velocities.

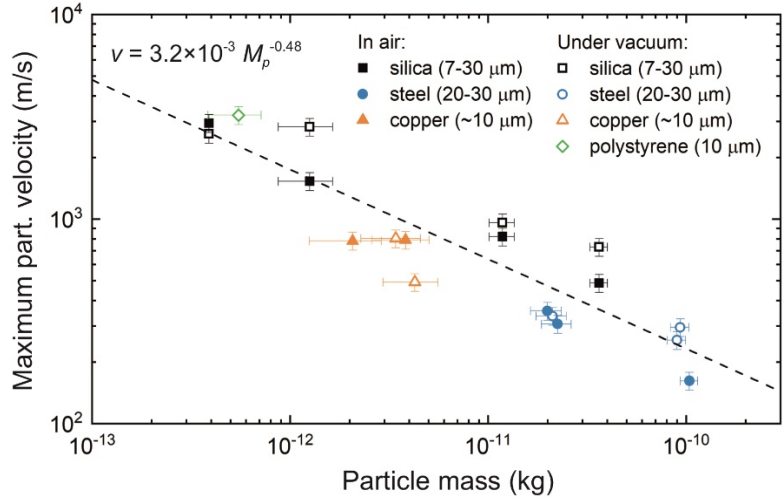

**Fig. 4.** Maximum particle velocity versus particle mass for a wide range of particle types and diameters under vacuum (0.3 mbar) and in atmospheric conditions.

## 3.2. Blast and drag considerations

Performing impact experiments under vacuum not only enables target testing under conditions relevant to space applications but also solves undesirable issues present in atmospheric conditions. The ablation process in air gives rise to an intense blast wave that can significantly affect target response and, in some cases, damage or destroy the target. In other work involving nanometric thin films (unpublished), we have repeatedly observed target bursting following blast exposure. As evidenced in Fig. 5a, the blast wave travels at supersonic velocities (in this case above 2 km/s) that can potentially overtake particles. In contrast, the blast is not observed under vacuum (Fig. 5b).

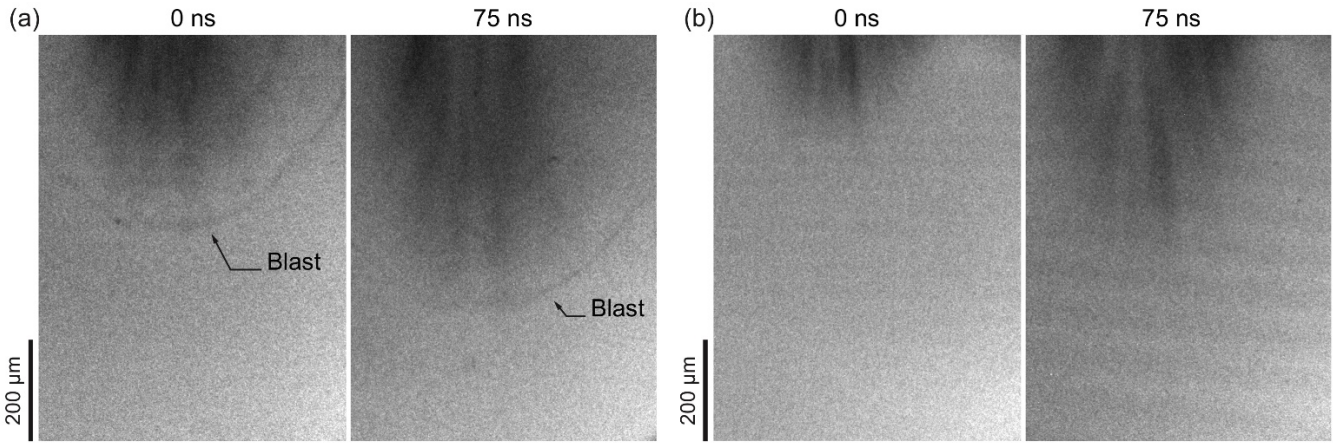

**Fig. 5.** (a) 2-frame sequence revealing blast wave and debris cloud generated upon laser ablation of gold film in atmospheric condition. (b) 2-frame sequence recorded under a vacuum of 0.28 mbar with the same laser excitation energy. Only the debris cloud is visible.

The blast wave can be tracked in real time and its intensity can be deduced from its velocity. In the example shown in Fig. 5a, we record the position of the front edge of the blast (Fig. 6a) and fit the trajectory $R_b(t)$ as a polynomial in order to calculate the blast velocity $v_b = dR_b/dt$. For a blast wave in air, the pressure is related to the blast velocity through the Mach number, $M$, with [26]:

$$\frac{P_b}{P_0} = \frac{2\gamma}{\gamma+1}(M^2 - 1), \qquad (4)$$

where $M = \frac{v_b}{c_0}$ and $\gamma = 1.4$. We calculate the corresponding blast-wave overpressure $P_b$ as a function of time (Fig. 6b). The overpressure is of the order of tens of bars, typical of laser-generated blast waves [27,28].

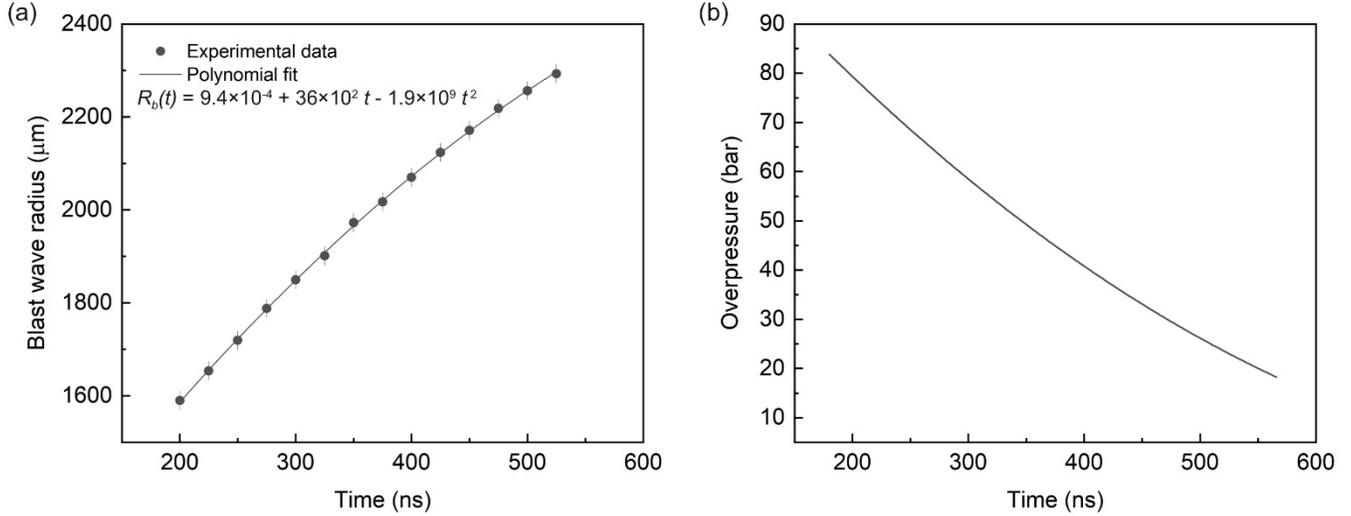

**Fig. 6.** (a) Blast wave position versus time and corresponding polynomial fit. (b) Calculated overpressure versus time.

For ambient conditions, one must also consider the drag exerted on a particle as it moves through air. Given the micrometer size of the particles, the velocity rapidly decreases on the μm-to-mm length scale because of the drag, which can be problematic when targets are placed far from the particle launch site.

Using hollow glass particles, whose low density enhances drag effects, we performed launch experiments to measure particle deceleration in air and in low vacuum. The particles were accelerated to about 270 m/s in air and 325 m/s in vacuum. Attempting to reach higher velocities by increasing the laser energy resulted in particle fracture due to their low strength. For these experiments, the field of view of the camera was centered about 1.5 mm away from launch site. We first notice that under the same laser excitation energy particles launched in air travelled at lower velocities than under vacuum. This is in part due to the drag exerted on the particles over the distance from the launch site to the camera field of view. The drag coefficient could be estimated from these observations; however, as the focused laser spot size varies with air pressure through the change of the refractive index of air, the particles were accelerated under different laser fluences, making direct comparison difficult. It is therefore preferable to directly observe particle trajectories in the camera field of view.

Considering the drag force acting on the particle, the equation of motion of the particle can be described as [29]:

$$M_p dv/dt = -1/2 \cdot C_D \rho_m A_p v^2. \qquad (5)$$

The density of air, $\rho_m$, is taken to be 1.2 kg/m$^3$ at ambient condition. The hollow particle has a diameter of 9 μm and a density of 670 kg/m$^3$. By integrating Eq. (5), the equation for the particle trajectory $x(t)$ as a function of time $t$ is:

$$x(t) = \frac{1}{B}\left[\ln\left(t - t_0 + \frac{1}{Bv_0}\right) - \ln\left(\frac{1}{Bv_0}\right)\right] + x_0. \quad (6)$$

with $v_0$ and $x_0$ the velocity and position at time $t_0$ and $B = (C_D \rho_m A)/(2M_p)$. Using the first recorded position of the particle as the reference time $t_0$ and the reference position $x_0$, the trajectories were extracted from the distortion-corrected videos and then fitted to find $C_D$ and $v_0$. Figure 7a shows the particle trajectory in air and reveals the deceleration of the particle due to drag. Repeating this measurement for 15 shots, we find a drag coefficient of $C_D = 0.7 \pm 0.2$. This value is in agreement with what is expected for a sphere moving through air at this velocity with a corresponding Reynolds number of ~200 [30]. This is the first direct experimental validation of this agreement at the micron scale.

In contrast, no deceleration of the particle was observed when launched under vacuum as shown in Fig. 7(b). The absence of deceleration under vacuum is particularly advantageous as the target can be placed farther away from the launch site, which allows more versatile target size ranges and geometries.

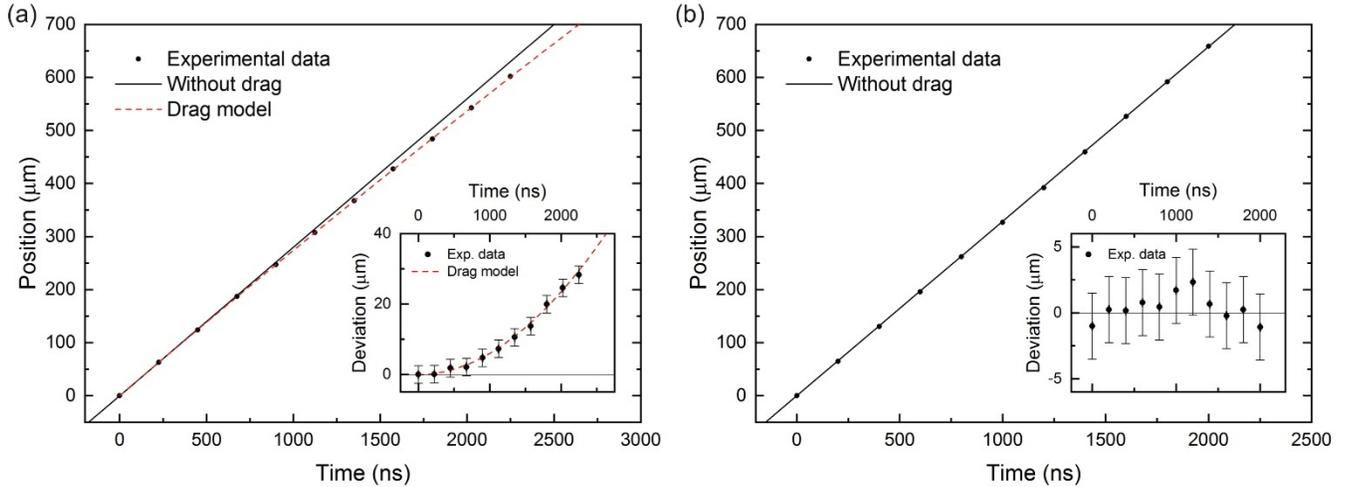

**Fig. 7.** (a) Hollow sphere trajectory under atmospheric conditions at $v_0 = 280$ m/s. The experimental particle trajectory (solid dots) is fitted following the drag model (red dotted line). For comparison, the black line represents the trajectory that the particle would have followed in the absence of drag. The deviation from this constant-velocity situation (drag-free situation) is shown in the inset. The error bars reflect the uncertainties in particle localization. (b) Hollow sphere trajectory under vacuum conditions (0.17 mbar) at $v_0 = 330$ m/s. The experimental trajectory is fitted using a linear regression. No deviation is observed from the constant-velocity trajectory, indicating negligible drag.

### 3.3. Particle impact on a polyethylene film

We demonstrate the material testing capabilities with microparticle impact of a polyethylene film. The film was made with 4000 average molecular weight polyethylene (purchased from Sigma-Aldrich). 1 gram of the material was placed on a glass slide and heated at 120°C until fully molten. Once fully molten, another glass slide was placed on top, separated by a 50-μm stainless steel spacer. Moderate pressure was applied to the top slide to ensure even thickness of the film. After 30 seconds, the glass slides were transferred to a beaker of water at room temperature for one minute. The glass slides were separated and the film was peeled off. The film was then allowed to air dry for one hour. The film was subsequently cut and tested in air in a free-standing configuration. The density of the film was 1200 kg/m³.

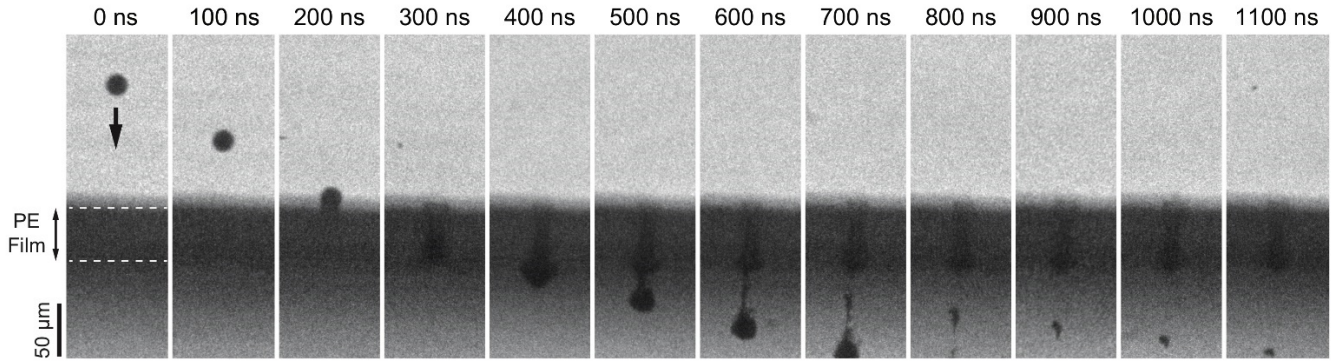

**Fig. 8.** Sequence of images showing a 19-μm steel particle impact on a polyethylene film at 520 m/s. Between frames 5 and 6, the particle exits the 50-μm film with a velocity of 265 m/s leaving an opened track in the sample behind it. 12 of 16 images are shown here cropped from their initial size for ease of visualization. The time stamps, shown above the frames, indicate the delay relative to the first frame of the sequence. The images were recorded using a 5-ns exposure time.

As shown in Fig. 8, a 19-μm steel particle ($\rho_p$ = 7800 kg/m³) impacts the film with a velocity of 520 m/s. In transparent materials, as in this case, we can observe the particle as it penetrates the film and opens a track in its wake. We see that upon exiting the film with a residual velocity of 265 m/s, the particle carries some of the film material that it encountered during penetration and additionally pulled material out behind it. We repeated such impacts for velocities from ~100 – 500 m/s and measured both impact and residual velocities, $v$ and $v_r$, respectively. We take $v_r$ to be negative for rebound and positive for perforation. At low velocities, we observed particle rebound up to a threshold velocity of about 230 m/s ($v_{th}$ in Fig. 9) above which particles remained embedded in the film. Above a second threshold velocity, namely the ballistic limit $v_{bl}$, we observed the particle exiting the film, similar to what is shown in Fig. 8.

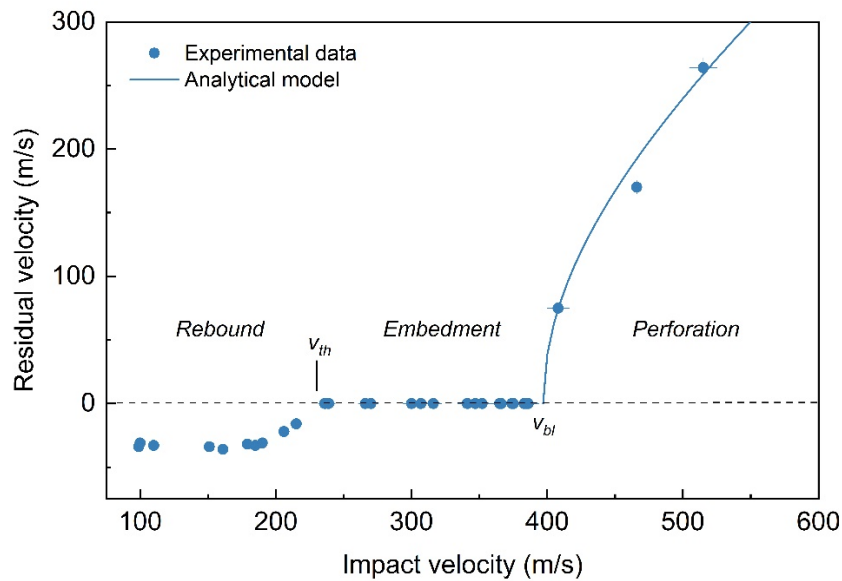

**Fig. 9.** Residual velocity versus impact velocity.

The model developed by Recht and Ipson [31] is often used to describe impacts resulting in perforation using simple energy considerations. Here, we consider, as a first approximation, that following penetration the projectile carries a plug of material, that remains attached to the projectile after it perforates the film. Through real-time observations (see Fig. 8), we notice that it is not exactly the case as we see small pieces of the specimen detached from the projectile. However, we also notice that these detached pieces are relatively small compared to the penetrated volume; therefore we neglect the kinetic energy contribution of the fragments compared to the kinetic energy of the projectile and the attached plug after perforation. We also assume no particle deformation. Applying conservation of energy, the kinetic energy of the incoming projectile equals the kinetic energy of the projectile and plug after perforation plus the energy $W$ associated with shearing, deformation, and heating of the plug and the penetrated region, giving:

$$\tfrac{1}{2} M_p v^2 = \tfrac{1}{2}(M_p + m_{pl}) v_r^2 + W. \qquad (7)$$

The energy loss due to deformation, heating, and shearing can be found using the velocity giving a residual velocity of zero (i.e. $v_{bl}$):

$$W = \tfrac{1}{2} M_p v_{bl}^2. \qquad (8)$$

For this polyethylene film, we find $W$ to be 2.2 µJ. The residual velocity can therefore be rewritten as:

$$v_r = \left( \frac{M_p}{M_p + m_{pl}} (v^2 - v_{bl}^2) \right)^{1/2}. \qquad (9)$$

This analytical solution is shown in Fig. 9 and matches reasonably well the experimental data points for perforation. As evidenced above, LIPIT accesses numerous interesting high-strain-rate material behaviors (rebound, embedment, shearing and perforation), captures those behaviors in real time, and enables quantitative analysis.

## 4. Conclusions

The LIPIT platform presented in this paper enables both the acceleration of a variety of microparticles to high velocities and the real-time visualization of single particle impacts on various materials. We demonstrated the current capabilities of the method, including particle velocity dependence with laser energy. We presented maximum velocities for a range of particles and discussed current limitations. The experimental setup can also be used to study multi-particle impact of fragmented and or molten particles. We showed that vacuum conditions reduce blast and drag. This platform can observe high-strain-rate material behavior such as failure and erosion and can facilitate the development of a wide range of materials to be exposed to high-velocity micro-particle impacts under both atmospheric and vacuum conditions. Finally, the present method allows access to a regime of particle sizes and velocities that is not attainable with conventional instruments.

## Acknowledgements

The authors declare no competing financial interests. The authors would like to thank Prof. Hajime Yano and his group members for fruitful discussions and for providing silica and polymeric particles. This material is based upon work supported in part by the U. S. Army Research Office through the Institute for Soldier Nanotechnologies, under Cooperative Agreement Number W911NF-18-2-0048.